\documentclass{iau}
\usepackage{graphicx}

\title[Physical and chemical properties of disks] 
{The physical and chemical properties of planet forming disks}

\author[Inga Kamp]   
{Inga Kamp$^1$}

\affiliation{$^1$Kapteyn Astronomical Institute, University of Groningen, PO Box 800, 9700 AV Groningen, \\ The Netherlands,  email: {\tt kamp@astro.rug.nl} \\[\affilskip]
}

\pubyear{2019}
\volume{345} 
\jname{Origins: from the Protosun to the First Steps of Life}
\editors{Bruce G. Elmegreen, L. Viktor T\'oth, Manuel G\"udel, eds.}
\begin{document}

\maketitle

\begin{abstract}
VLT instruments and ALMA have revolutionized in the past five years our view and understanding of how disks turn into planetary systems. They provide exquisite insights into non-axisymmetric structures likely closely related to ongoing planet formation processes. The following cannot be a complete review of the physical and chemical properties of disks; instead I focus on a few selected aspects. I will review our current understanding of the physical properties (e.g. solid and gas mass content, snow and ice lines) and chemical composition of planet forming disks at ages of 1-few Myr, especially in the context of the planetary systems that are forming inside them. I will highlight recent advances achieved by means of consistent multi-wavelength studies of gas AND dust in protoplanetary disks.
\end{abstract}

\firstsection
\section{Introduction}

The physical properties have important consequences for the formation of planets
such as how much 'solid' surface density is available in the various planet
forming regions, the gravitational stability of disks, as well as the total disk
gas mass available to keep planet formation ongoing at a few Myr. I will review
here the icelines in disks --- both from the modeling perspective as well as
recent observational insights --- and gas masses, mostly summarizing the current
caveats and debates on the interpretation of CO rotational line observations and
the prospects for using HD instead in future missions.

The chemical properties have implications for the composition of forming planets.
The disk composition determines the composition of planetsimals and hence that of
growing planetary cores, e.g.\ the water content as a function of distance from
the star and also the C/O ratio of planets and the composition of
primary/secondary atmospheres. Two topics that received quite some attention in
the past few years are the aspect of heritage versus in-situ for the chemical
composition of planet forming disks and the role of (warm) surface chemistry for
the 'solid' composition, ice and refractory material among them phyllosilicates;
I focus here on these two.

These selected topics all demonstrate the close coupling between gas and dust in
planet forming disks and its high relevance for the interpretation of
observations. This coupling happens not only thermally through collisions, but
also through grain opacities (e.g.\ non-thermal desorption processes), IR pumping
for molecular excitation, gas pressure determining dust grain settling and
migration, and varying effective grain surface area for chemical reactions
throughout the disk. A proper interpretation of the rich existing and upcoming
disk observations requires state-of-the-art disk models to include these coupling
terms.

\section{Disk structure}

Our knowledge of disk structure is largely based on seminal papers of steady
state accretion disks (\cite[Pringle 1981]{Pringle1981}) and irradiated disks
(\cite[Kenyon \& Hartmann 1987]{Kenyon1987}; \cite[Chiang \& Goldreich
1997]{Chiang1997}). Gas and dust temperatures decouple in the disk surface
(\cite[Kamp \& Dullemond 2004]{Kamp2004}; \cite[Jonkheid et al.
2004]{Jonkheid2004}). While dust is mainly heated by the bulk luminosity of the
central star, the gas thermal balance is driven by far-UV (FUV) and X-rays from
the central star, which are absorbed and scattered by dust grains in the disk
surface (see \cite[Dullemond et al. 2007]{Dullemond2007}, for a review of disk
structure). X-rays mainly interact with the gas that similarly absorbs and
scatters them (\cite[Nomura et al. 2007]{Nomura2007}; \cite[Aresu et al.
2011]{Aresu2011}); if grains have grown much larger than the ISM, dust opacities
become only important for high energy X-rays ($E\gtrsim 5-10$~eV, \cite[Rab et
al. 2018]{Rab2018}).

\begin{figure}[t]
\begin{center}
\includegraphics[width=12cm]{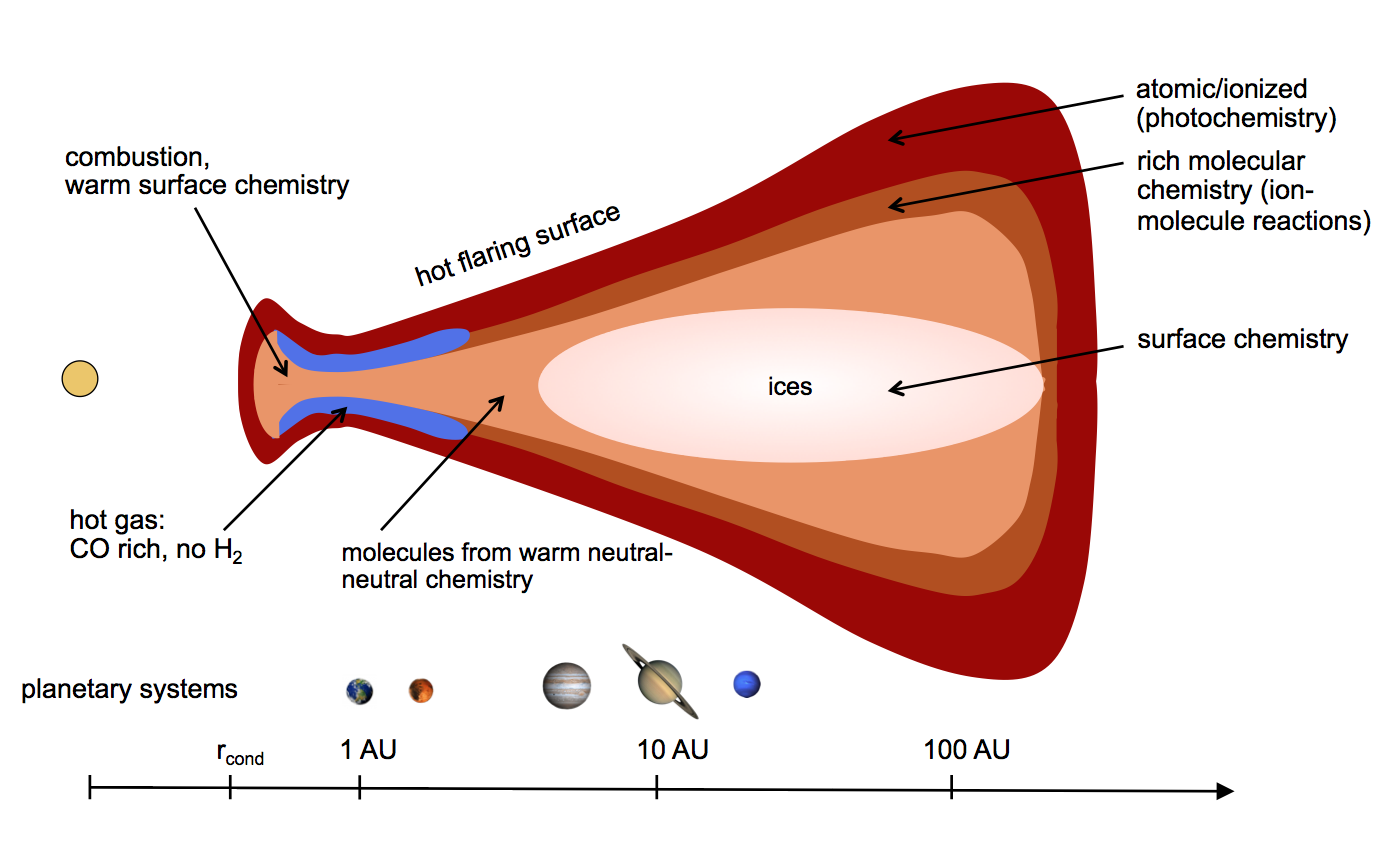}
\end{center}
\vspace*{-3mm}
\caption{Chemical structure of a planet forming disk outlining the layered
structure (atomic/molecular/ice) and the various types of chemistry
(photochemistry, ion-molecule chemistry, neutral-neutral chemistry,
combustion chemistry) relevant in the different parts of the disk.}
\label{fig-chemical-layering}
\end{figure}

Cosmic rays (CR) are less relevant for heating the gas, but they drive chemical
processes close to the disk midplane such as non-thermal desorption and cosmic
ray ionization. Disks are layered structures (e.g. \cite[Aikawa et al.
2002]{Aikawa2002}; \cite[Kamp \& Dullemond 2004]{Kamp2004}, see
Fig.~\ref{fig-chemical-layering}) that can be divided into
\\[-3mm]
\begin{enumerate}
\item a hot ionized/atomic layer in which photochemistry dominates \\[-3mm]
\item a warm molecular layer in which ion-molecule chemistry dominates \\[-3mm]
\item a cold icy midplane in which CR ionization and surface chemistry dominate.\\[-3mm]
\end{enumerate}
Surface chemistry encompasses a wide range of non-thermal desorption processes
and has been reviewed recently by \cite[Cuppen et al. (2017)]{Cuppen2017}. In
addition to these three disk layers, \cite[Lee et al. (2010)]{Lee2010} discussed
implications of combustion chemistry for the innermost parts of protoplanetary
disks. A more general review of chemical evolution in disks including
observations can be found in \cite[Bergin et al. (2007)]{Bergin2007} and
\cite[Dutrey et al. (2014)]{Dutrey2014}.

Planet forming disks are shaped by dynamical processes such as accretion onto the
central star, winds (radiation or magnetically driven), gas turbulence, vertical
settling, collisional growth and radial migration of dust grains, as well as
disk-planet interaction (e.g. \cite[Weidenschilling 1977]{Weidenschilling1977};
\cite[Miyake \& Nakagawa 1995]{Miyake1995}; \cite[Dullemond \& Dominik
2004]{Dullemond2004}). \cite[Alexander (2008)]{Alexander2008}, \cite[Armitage
(2011)]{Armitage2011} and \cite[Birnstiel et al. (2016)]{Birnstiel2016} provide
recent reviews of the various processes.

Over the past decade, there has been increasing observational evidence that
confirms the general picture described above. \cite[Avenhaus et al.
(2018)]{Avenhaus2018} found in their VLT/SPHERE survey of T~Tauri disks that the
scattered light surface displays a typical flaring structure; previously this had
been traced with resolved PAH emission for the Herbig disk around HD\,97048
(\cite[Lagage et al. 2006]{Lagage2006}). In addition, dust settling has now been
directly seen with ALMA. \cite[Mulders \& Dominik (2012)]{Mulders2012} show that
in a self-consistent settling model, mm-sized dust grains are concentrated at
heights of $z/r \lesssim 0.02$. Modeling the high spatial resolution HL~Tau ALMA
data (\cite[ALMA Partnership et al. 2015]{ALMA2015}), \cite[Pinte et al.
(2016)]{Pinte2016} demonstrate that the similarity in gap width between major-
and minor-axis requires a typical $z/r$ of $\lesssim 0.01$, consistent with a
very low turbulence, $\alpha=3\times10^{-4}$.

\begin{figure}[h]
\begin{center}
\includegraphics[width=11cm]{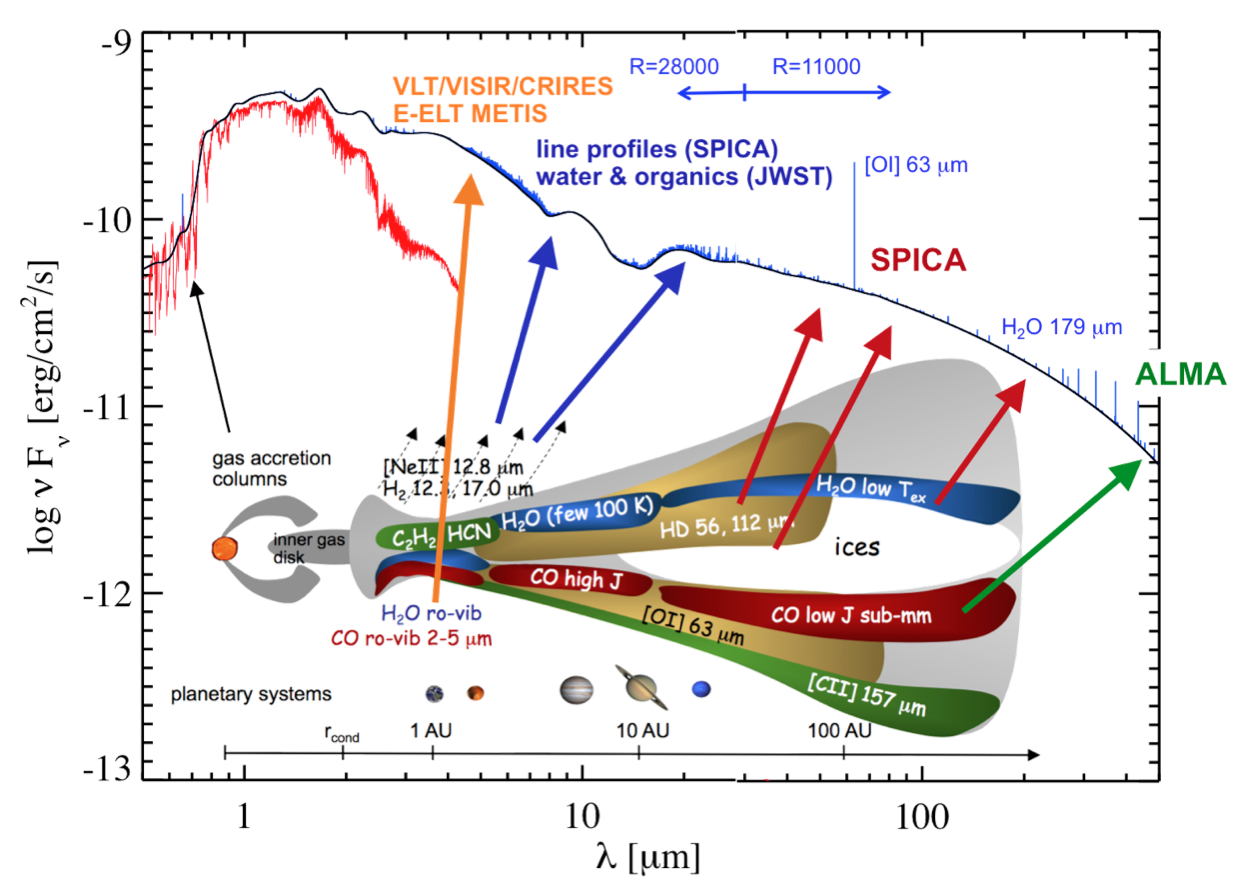}
\end{center}
\vspace*{-2mm}
 \caption{An overview of line emission from near-IR to submm wavelengths from planet
 forming disks and existing and upcoming instrumentation to detect it. The model
 SED is that of a typical T~Tauri disk using two different spectral resolutions
 ($R=28\,000$ and $R=11\,000$ relevant for the planned SMI and SAFARI instruments
  onboard the proposed SPICA mission).}
   \label{fig-disk-line-emission}
\end{figure}

Understanding the physical conditions of the gas in these planet forming disks is
challenging because gas lines emit according to their respective excitation
conditions which depend to first order on the gas temperature, the density and
the optical depth. Figure~\ref{fig-disk-line-emission} illustrates how emission
from hot gas in the inner disk occurs preferentially at near- to mid-IR
wavelengths. For these lines, the dust optical depth at the corresponding
wavelength determines how deep we can peer into these disks. Molecules such as CO
and water, which possess a wide range of lines across the near- to far-IR
wavelength range, can scan the surface and also probe successively deeper into
the disk as the dust becomes more optically thin towards longer wavelengths.
Since each line probes only a small region in such a planet forming disk, we need
a suite of instruments to trace the gas across the entire disk such as VLT/VISIR
and CRIRES (ELT/METIS in the future) for the hot (few 100 - few 1000~K) gas at
short wavelengths, JWST/NIRSPEC and MIRI in the near- to mid-IR, the proposed
SPICA mission ranging from mid- to far-IR and bridging the gap to detecting cold
(few 10~K) gas with ALMA at submm wavelengths.

We are now in an era where we have tens of disks with very rich multi-wavelengths
and multi-kind observational data: X-ray to mm wavelengths and photometric,
spectroscopic, spectro-astrometric, imaging and interferometric datasets for both
gas and dust. Disk modeling has caught up with the challenge to interpret such
complex datasets in the form of various radiation thermo-chemical models such as
{\sc ProDiMo} (\cite[Woitke et al. 2009]{Woitke2009}; \cite[Kamp et al.
2010]{Kamp2010}; \cite[Thi et al. 2011]{Thi2011a}), {\sc DALI} (\cite[Bruderer et
al. 2012]{Bruderer2012}), and {\sc ANDES} (\cite[Akimkin et al.
2013]{Akimkin2013}). The DIANA project has recently demonstrated for a sample of
14 objects that individual self-consistent parametrized dust+gas disk models are
able to match the entirety of such complex datasets (\cite[Woitke et al.
2018]{Woitke2018}, e.g. line fluxes across all wavelengths match observed ones to
within a factor three). Such modeling can break the large degeneracies in the
interpretation that often exist when only individual datasets are analysed.

\section{Disk physical properties}

These properties provide important constraints for planet formation models. The
following paragraphs present two examples that received a lot of attention over
the last few years, especially in the context of new ALMA observations: The
location of icelines in disks and the determination of gas masses.

\subsection{Icelines}

Icelines in disks are not sharp transitions. Different binding energies for water
on bare and icy grains as well as turbulent mixing lead to so-called 'snow
borders' in which bare grains and icy grains coexist. \cite[Marseille \& Cazaux
(2011)]{Marseille2011} explore in their models the radial width of such 'snow
borders' and find typical values of $0.03-0.2$~au. Using a completely different
modeling approach, \cite[Krijt et al. (2016)]{Krijt2016} explore the vertical
distribution of water vapor and ice in the presence of dust grain growth,
settling and turbulent mixing. At low turbulence levels ($\alpha \leq 10^{-3}$),
the snowline resides about one pressure scale hight deeper in the disk compared
to simple equilibrium models. \cite[Ciesla \& Cuzzi (2006)]{Ciesla2006} and
\cite[Furuya \& Aikawa (2014)]{Furuya2014} showed that --- besides relocating
snowlines
--- turbulent mixing and transport processes have large effects on the chemical
composition of disks, especially the local abundances of water, ices and complex
organic molecules. It is however notable that for example vertical column
densities of gaseous CO in the outer disk ($\gtrsim 50$~au) are hardly affected
by turbulent mixing; this confirms earlier results by \cite[Willacy et al.
(2006)]{Willacy2006}.

At lower temperatures ($<100$~K), in the outer disk, photodesorption plays an
important role in setting the height of the icelines. \cite[Ceccarelli \& Dominik
(2005)]{Ceccarelli2005} postulated this effect to explain the high level of
deuterated cold water emission in DM~Tau; later Herschel/HIFI observations
detected this "photodesorbed" water reservoir through the ground state lines of
the main isotopologue H$_2$O (\cite[Hogerheijde et al. 2011]{Hogerheijde2011};
\cite[Podio et al. 2013]{Podio2013}). Laboratory and thermo-chemical disk
modeling efforts subsequently quantified the process (e.g. \cite[Fraser et al.
2011]{Fraser2001}; \cite[\"Oberg et al. 2009]{Oeberg2009}; \cite[Kamp et al.
2013]{Kamp2013}).

Very recently, \cite[Schwarz et al. (2016)]{Schwarz2016} and \cite[Pinte et al.
(2018)]{Pinte2018} directly observed the effect of photodesorption using ALMA CO
isotopologue observations in the disks of TW~Hya and IM~Lup respectively. The CO
temperature plateaus at 21~K at intermediate radii in both datasets; this is
consistent with laboratory measurements of CO binding energies. In addition, the
radial extent and location of CO as derived from the high spatial resolution
channel maps can only be reproduced with models that take into account CO
photodesorption (\cite[Pinte et al. 2018]{Pinte2018}). The depth to which photons
can penetrate these disks and drive photodesorption depends strongly on the local
dust properties, demonstrating once more the high relevance of consistent dust
and gas models in interpreting the new data.

\subsection{Gas masses}

For decades, many continuum surveys measured disk masses as a key input for the
"planet forming potential" of these disks (e.g. \cite[Beckwith et al.
1990]{Beckwith1990}; \cite[Andre \& Montmerle 1994]{Andre1994}; \cite[Andrews \&
Williams 2005]{Andrews2005}; \cite[Andrews \& Williams 2007]{Andrews2007}).
Photometric measurements at long wavelength, where the emission is thought to be
optically thin, have been used together with a representative dust emissivity
(opacity) to quantify disk dust masses\footnote{Note that dust mass refers here
to the solid mass in small dust grains up to mm-sizes; the solid mass in larger
planetesimals and even protoplanets does not contribute significantly to the mm
fluxes of disks and can hence not be measured this way.}. To probe how disks
evolve into planetary systems, the gas-to-dust mass ratio was often taken as an
indicator, primordial disks being rich in gas with a gas-to-dust ratio of $\sim
100$ and debris disks being gas-poor. Several gas surveys (e.g.\ FEPS, GASPS)
aimed at deriving this key quantity for various star forming regions using
tracers such as mid-IR line emission (warm gas) and CO submm lines (cold gas) and
the [O\,{\sc i}]\,63~$\mu$m fine structure line (e.g. \cite[Pascucci et al.
2006]{Pascucci2006}; \cite[Dent et al. 2013]{Dent2013}).

With the advent of more sensitive submm interferometers such as SMA and ALMA,
\cite[Williams \& Best (2014)]{Williams2014} suggested to use the two rarer
isotopologues $^{13}$CO and C$^{18}$O to determine cold gas masses in disks
(diagnostic CO isotopologue mass diagram) and found gas-to-dust ratios below 20
in several Taurus disks. The work of \cite[Willacy et al. (2006)]{Willacy2006}
and \cite[Furuya \& Ailawa (2014)]{Furuya2014} had shown that low levels of
turbulence as detected in several planet forming disks (e.g. \cite[Flaherty et
al. 2018]{Flaherty2018}, for TW~Hya) should not affect the CO abundance in the
outer disk. \cite[Miotello et al. (2016)]{Miotello2016} show that isotope
selective photodissociation can affect the gas measurements by an order of
magnitude. Still ALMA surveys report gas-to-dust mass ratios well below 100 in
1-3~Myr old star forming regions (e.g. \cite[Miotello et al. 2017]{Miotello2017},
for Lupus). Recently, \cite[Woitke et al. (2016)]{Woitke2016} and \cite[Kamp et
al. (2018)]{Kamp2018} quantify the effects of changing dust properties and disk
outer edges (tapering off) for the CO isotopologue ratio and the diagnostic CO
isotopologue mass diagram. The DIANA multi-wavelength disk modeling project has
shown that a sample of ten well known disks span over two orders of magnitude in
CO isotopologue line fluxes, while the gas-to-dust mass ratio in the
thermo-chemical dust+gas models, that are consistent with an even wider
observational dataset, span gas-to-dust mass ratios between 50 and 450
(\cite[Kamp et al. 2018]{Kamp2018}). It is important to note that the "main
sequence" of CO isotopologue lines is consistently found in all disk modeling
works (\cite[Williams \& Best 2014]{Williams2014}; \cite[Miotello et al.
2016]{Miotello2016}; \cite[Kamp et al. 2018]{Kamp2018}), but disk mass is not the
only parameter shifting line fluxes up and down this main sequence; for example
the maximum grain size or the power law index of the grain size distribution also
play a key role. Using the now available multi-band medium resolution ALMA
surveys of several star forming regions, the underlying dust properties can be
constrained observationally and consistent gas+dust modeling should be employed
to derive better estimates of gas masses from CO isotopologue data. However, the
statistics are often limited since many of the surveys are fairly shallow and the
number of sources detected in both $^{13}$CO and C$^{18}$O remains small compared
to the total number of sources detected in the dust continuum.

Since the HD abundance depends only on the primordial D abundance, an alternative
to determine warm gas masses is the use of the two lowest excitation HD lines at
112 and 56~$\mu$m ($T_{\rm ex}=128.5$ and $384.3$~K respectively). \cite[Bergin
et al. (2013)]{Bergin2013} and later \cite[McClure et al. (2016)]{McClure2016}
report HD masses or upper limits for seven T Tauri disks. Interestingly, the gas
masses derived from the warm HD lines via disk modeling are often larger than
those derived via the CO lines. \cite[Trapman et al. (2017)]{Trapman2017} use
thermo-chemical disk models to explore the predictive power of these HD lines for
disk gas masses and show that deep HD surveys with the proposed JAXA/ESA SPICA
mission can detect disk gas masses down to $10^{-5}$~M$_\odot$ for typical T
Tauri disks in nearby low mass star forming regions.

\section{Disk chemical properties}

These properties provide constraints for the composition of the planets forming
within these disks. The following paragraphs review two controversial issues
relating to these chemical properties: The question of heritage versus in-situ
for the chemical composition of disks and the relevance of surface chemistry in
disks.

\subsection{Heritage versus in-situ}

The thermal history of the material forming a disk is strongly affected by
processes such as stellar outbursts, and grain growth during the collapse/disk
formation phase. The collapse of a dense core is dynamically complex involving
magnetic fields and possibly instabilities. A showcase example of this complexity
is the ALMA high spatial resolution imaging of the binary protostar system IRAS
16293-2422 (PILS, \cite[J{\o}rgensen et al. 2016]{Joergensen2016}).

\begin{figure}[h]
\begin{center}
\includegraphics[width=12.cm]{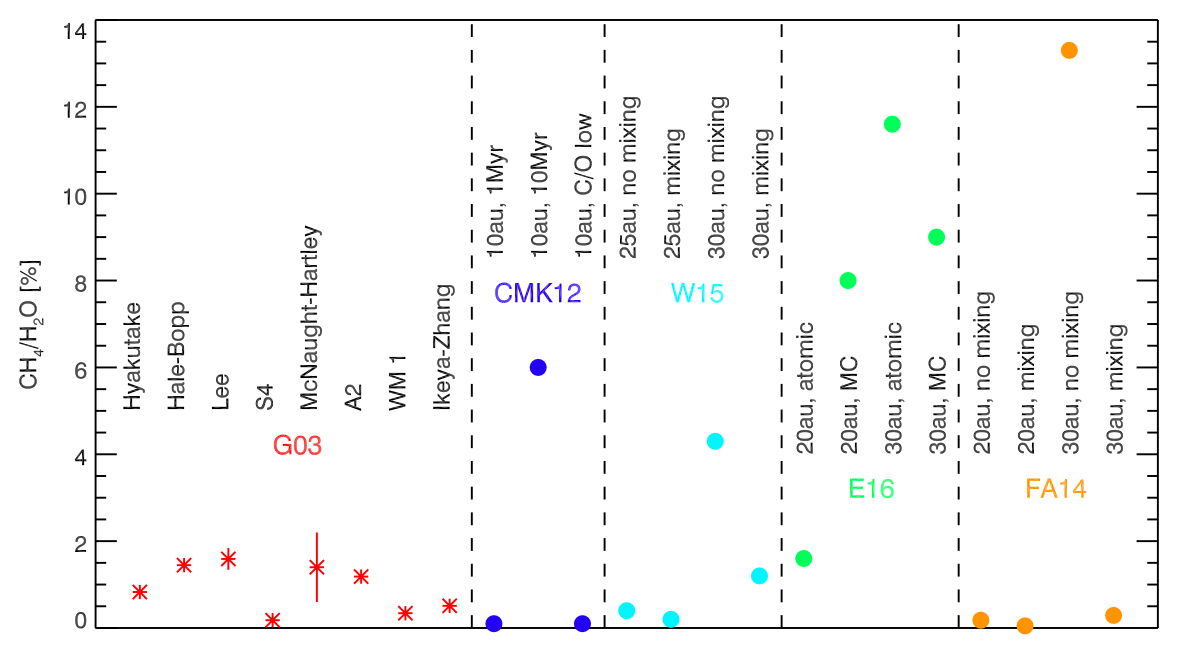}
\end{center}
{\caption{Comparsion between a collection of methane/water ice
abundance ratios (with error bars) in Oort
cloud comets (\cite[Gibb et al. 2003]{Gibb2003}) and a sample of disk models:
\cite[Chaparro et al. (2012, blue, C/O$=0.45$, low C/O$=0.16$)]{Chaparro2012},
\cite[Willacy et al. (2015, cyan)]{Willacy2015}, \cite[Eistrup et al. (2016,
green)]{Eistrup2016}, and \cite[Furuya \& Aikawa (2014, orange)]{Furuya2014}.
Figure updated from a previous version by E.~Gibb
(private communication).}\label{fig-comets}}
\vspace*{-2mm}
\end{figure}

Several works studied the ice processing during the disk formation phase (e.g.
\cite[Chick \& Cassen 1997]{Chick1997}; \cite[Visser et al. 2009] {Visser2009};
\cite[Visser et al. 2011]{Visser2011}; \cite[Furuya et al. 2012]{Furuya2012};
\cite[Drozdovskaya et al. 2014]{Drozdovskaya2014}). The comparison of comets and
outer disks composition can provide important clues as to the amount of
additional processing within the disk and the thermal history of ices.
\cite[Willacy et al. (2015)]{Willacy2015} show that most (D/H)$_{\rm water}$
measurements are only compatible with disk models that include mixing processes.
\cite[Cleeves et al. (2014)]{Cleeves2014} go one step further using a
comprehensive range of disk ionisation sources and a detailed chemical model of
the protosolar disk to claim that a significant fraction of HDO in the disk stems
from accretion and inheritance of interstellar ices. Figure~\ref{fig-comets}
shows a comparison between the methane/water abundance ratio in cometary ices and
the same ratio in a series of disk models. It becomes clear that this ice ratio
depends strongly on the initial condition (atomic versus molecular), the
elemental abundances (not only C and O, but also other metals as electron
sources), mixing, as well as the chemical age of the disk. The no-mixing models
and the young chemical ages give a better agreement with the observed
CH$_4$/H$_2$O ratio than mixing/old models (contrary to D/H). The initial
composition (atomic or molecular cloud values) as well as the element abundances
(C/O) play a large role in setting the ice abundances in the outer disk (beyond
10~au). However, we are only at the beginning since many other effects also
impact the ice composition, especially also the relative abundances of ices, such
as layering and diffusion within ices, grain growth and migration, and the role
of irradiation within ices.

\subsection{(Warm) Surface chemistry}

Surface chemistry itself is less well understood compared to gas-phase chemistry
and the status and current limitations are comprehensively reviewed by
\cite[Cuppen et al. (2017)]{Cuppen2017}. Surface chemistry is often invoked to
explain the formation of COMs (complex organic molecules with six or more atoms,
at least two of which are C, N, O). \cite[Ciesla \& Sandford (2012)]{Ciesla2012}
form COMs via mixing of icy grains in disks, subsequently exposing those icy
grains to different levels of UV dosage. \cite[Drozdovskaya et al.
(2016)]{Drozdovskaya2016} propose that the abundances of COMs are strongly
affected by the dynamical history of the disk build-up evolution (pure infall
versus infall+viscous spreading). \cite[Walsh et al. (2014)]{Walsh2014} show that
even without taking details of disk build-up into account, significant abundances
of COMs can be synthesized in disks through grain surface chemistry.

But also warm surface chemistry has recently received more attention in studying
the composition of the inner few au as well and the link to terrestrial planets
and asteroids. Combustion chemistry can burn carbonaceous dust inside $\sim 1$~au
(e.g. \cite[Lee et al. 2010]{Lee2010}; \cite[Anderson et al.
2017]{Anderson2017}), although the level of carbon depletion may be limited by
the efficiency of radial and vertical mixing processes, which replenish carbon
rich material faster than burning can proceed (\cite[Klarmann et al.
2018]{Klarmann2018}).

A second process driven by warm surface chemistry is the transformation of
silicates into phyllosilicates (sheet silicates with intercalated water). Early
work by \cite[Fegley \& Prinn (1989)]{Fegley1989} dismissed the relevance of this
process based on the long timescales it required compared to the lifetime of
disks. More recent works by \cite[Stimpfl et al. (2006)]{Stimpfl2006} and
\cite[Muralidharan et al. (2008)]{Muralidharan2008} using realistic silicate
surfaces and adsorption energies find higher efficiencies in the solar nebula
around the region where Earth has formed. We recently revisited these processes
using both Monte Carlo simulations to assess the efficiency of water adsorption
under high temperature/pressure conditions inside $\sim 1$~au (\cite[D'Angelo et
al. 2018]{DAngelo2018}) and the subsequent surface chemistry that transforms part
of the silicate grains into phyllosilcate material (\cite[Thi et al.
2018)]{Thi2018}. Inside the water snowline, this process can proceed within 1~Myr
to produce grains that contain a significant fraction of water (up to a few \%).
If the process occurs before grains have substantially grown, this is an
alternative pathway to incorporate water into the proto-Earth, i.e.\ contribute
to the water content within the Earth mantle.

\section{Outlook}


The future has great prospects to employ disk observations in conjunction with detailed disk models (radiation, physics, chemistry, dynamics) to achieve a more comprehensive view on planet formation. ALMA and VLT (now and in the future) will provide molecular channel maps to confront thermo-chemical disk models (gas+dust), to characterize rings and non-axisymmetric gas structures (embedded protoplanets, vortices), to detect circumplanetary disks, and study inner disk dispersal processes. In the near future, SOFIA/HIRMES ($\sim 2019$) will detect thermal ice features and HD in a about a dozen disks showcasing the potential of such data. From 2021 onwards, JWST will provide data to study the warm gas+dust composition/evolution in the inner $\sim 10$~au. ELT/METIS (beyond $\sim 2025$) will provide detailed kinematics (channel maps) in the inner $\sim 10$~au of disks, and study the formation and properties of circumplanetary disks. Further ahead beyond 2030, the proposed SPICA mission can study the trail of water, ice and solids during the entire planet formation phase through large unbiased surveys of water, HD, ices, and dust mineralogy well into the debris disk stage.


\begin{discussion}

\discuss{Marov}{What mechanism of dust particles integration do you postulate in your model --- whether you believe it could be by integration per se through a direct collision mechanism, or rather some interaction process involving, say, original dust cluster formation is to be invoked? In my mind the latter would significantly ease the integration and particle growth process, which is supported by the results of my team computer modeling. }

\discuss{Kamp}{First of all, I myself have no model of dust particle growth. Growth is now already detected at very early stages of star formation, e.g. coreshine. The very first condensation of dust is indeed thought to proceed via clusters that form e.g.\ around AGB stars. However, similar processes can occur if refractory material sublimates in disks close to the star due to e.g.\ outbursts and then recondenses.}

\end{discussion}

\end{document}